\documentclass[prl,twocolumn,nofootinbib,superscriptaddress]{revtex4-2}
\pdfoutput=1

\usepackage{amsfonts}
\usepackage{amsmath}
\usepackage{amssymb}
\usepackage{amsthm}
\usepackage{bm}
\usepackage{hyperref}

\usepackage{dcolumn}
\usepackage{epsfig}
\usepackage{graphicx}
\usepackage{graphics}
\usepackage{latexsym}
\usepackage{rotating}
\usepackage[lined,boxed]{algorithm2e}

\begin{document}

\title{Non-Convex Optimization by Hamiltonian Alternation}

\author{Anuj Apte}
\affiliation{Department of Physics, University of Chicago, Chicago, IL 60637}
\author{Kunal Marwaha}
\affiliation{Department of Computer Science, University of Chicago, Chicago, IL 60637}
\author{Arvind Murugan}
\email{amurugan@uchicago.edu}
\affiliation{Department of Physics, University of Chicago, Chicago, IL 60637}

\begin{abstract}
    A major obstacle to non-convex optimization is the problem of getting stuck in local minima.
    We introduce a novel metaheuristic to handle this issue, creating an alternate Hamiltonian that shares minima with the original Hamiltonian only within a chosen energy range. 
    We find that repeatedly minimizing each Hamiltonian in sequence allows an algorithm to escape local minima.
    This technique is particularly straightforward when the ground state energy is known, and one obtains an improvement even without this knowledge.
    We demonstrate this technique by using it to find the ground state for instances of a Sherrington-Kirkpatrick spin glass.
\end{abstract}

{
\let\clearpage\relax
\maketitle
}

\section{Introduction}
A \textit{non-convex} optimization problem contains a constraint or cost function that is not convex.
These problems are relevant to fields as diverse as the theory of spin glasses \cite{Fyodorov_2004, Cerny2010}, computational chemistry \cite{Floudas_Pardalos_2011}, machine learning \cite{Jain_Kar_2017}, and operations research \cite{Hiriart-Urruty_1989}.
As a result, non-convex optimization is a subject of considerable theoretical and practical significance. 
In a \emph{convex} optimization problem, all locally optimal solutions are globally optimal \cite{Bertsekas_2009}; in contrast, non-convex optimization has no such guarantee.
In general, checking whether a solution is the global minimum of a non-convex optimization problem is NP-complete~\cite{Murty_Kabadi_1987}, and several non-convex problems are NP-hard to approximately solve \cite{Meka_2008}. 

Practitioners often use quasi-Newton methods~\cite{Nocedal_Wright_2006} and stochastic gradient descent~\cite{Rustagi_1971} to approximately solve non-convex optimization problems. These algorithms provably converge to a local minimum.
However, the global performance of these algorithms is not well-understood: non-convex cost functions (or Hamiltonians\footnote{
In this paper, we use ``Hamiltonian'' and ``cost function'' interchangeably. In physics and chemistry, the cost function is typically the Hamiltonian of the system under study.})
may have several local minima with energy well above the global minimum.
Any effective algorithm for non-convex optimization must grapple with the problem of getting stuck in these local minima.

How does an algorithm escape local minima? 
One popular strategy, named \emph{multi-start}, is to run several copies of the algorithm, each with a different starting point.
These iterations increase the chance that at least one copy converges to the global minimum~\cite{Kocsis_2011}. 
Another strategy, often inspired by intuition about physical dynamics, allows the minimization algorithm to take steps that do not necessarily minimize the function. An example of such an algorithm is simulated annealing, where the probability of taking a non-improving step is set by an ``annealing  schedule'' that changes the effective ``temperature'' of the algorithm's search dynamics~\cite{Kirkpatrick_Gelatt_Vecchi_1983}. This concept can be generalized by incorporating random external fields to drive non-improving steps~\cite{Zarnd2002}. 

A different approach to escape local minima uses time-varying landscapes, switching between the cost function to be optimized and other functions with correlated features. For example, biological evolution in time-varying fitness landscapes has been proposed \cite{Alon_2005,Sachdeva_2020,Murugan_2021} as a way to drive evolution towards otherwise inaccessible states. Related ideas of cyclic landscape protocols have been explored in computational chemistry \cite{stillinger1988nonlinear} and glass physics \cite{keim2014mechanical,pine2005chaos}. 
In the context of machine learning, these protocols \cite{ruiz-garcia21} are connected to concepts behind curriculum learning \cite{bengio2009curriculum}.
Note that the alternating landscapes in these works are not freely chosen.
Furthermore, the effectiveness of cycling is only understood at a heuristic level using
statistical correlations between extrema of the alternating landscapes \cite{Sachdeva_2020}. 

\begin{figure}[htb!]
\includegraphics[width=0.35\textwidth,height= 0.25\textwidth]{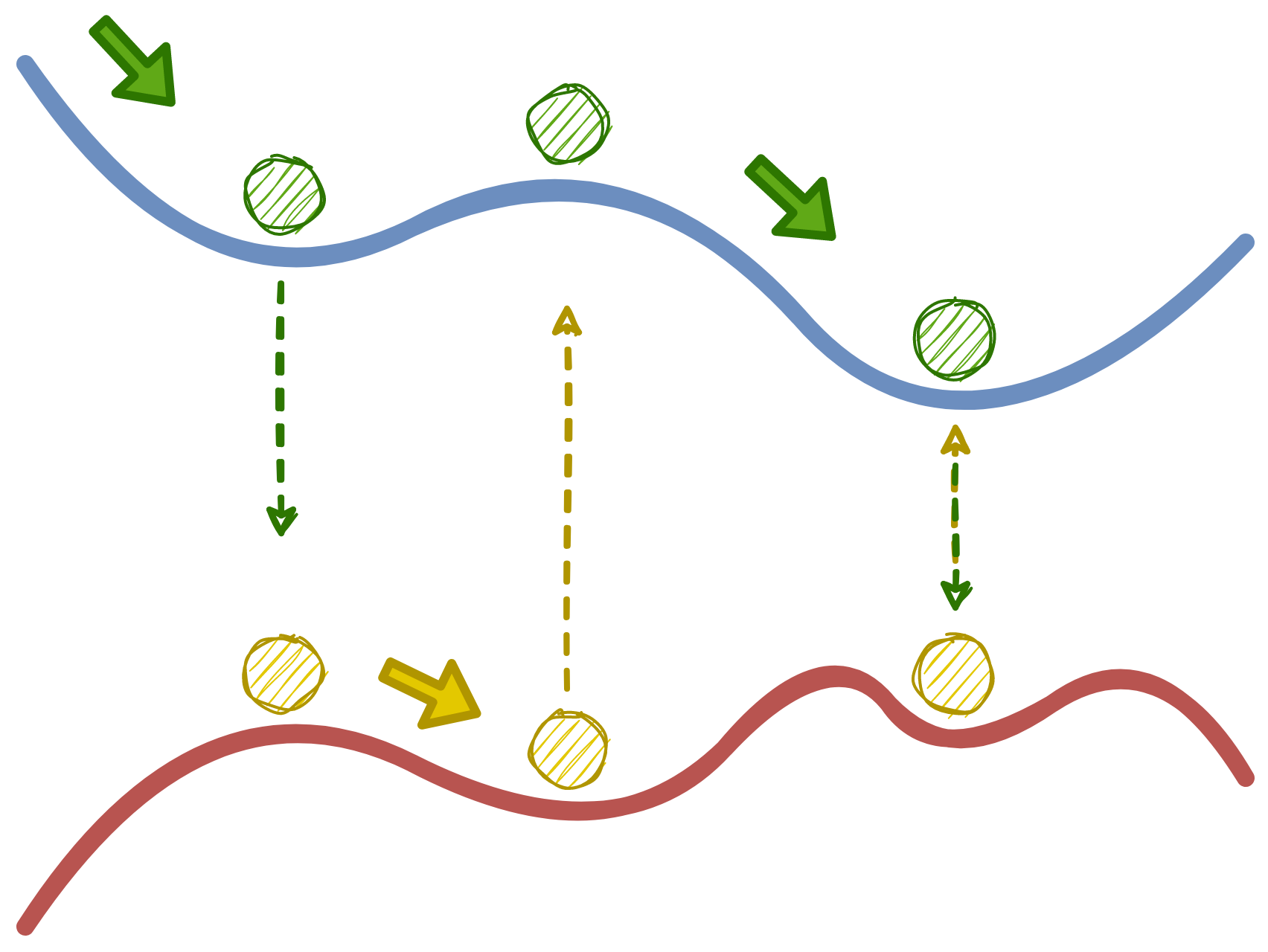}
    \caption{\footnotesize
    A schematic illustrating the intuition behind the \emph{Hamiltonian alternation} strategy.
    A particle initially moves in an energy landscape (blue) defined by a Hamiltonian $H(x)$. We analytically construct a second Hamiltonian $\widetilde{H}(x) = K(H(x))$ (red) such that all local minima of $H(x)$ above a threshold energy are guaranteed to be local maxima of $\widetilde{H}(x)$. 
    Switching particle dynamics between the two energy landscapes (red and blue) allows the particle to escape local minima of $H(x)$. However, since the ground state of $H(x)$ is also a minimum of $\widetilde{H}(x)$ (see rightmost minimum), the ground state of $H(x)$ is an attractor of the switching dynamics. Once the particle has settled here, switching landscapes will not change the location of the particle.}
\label{cartoon}
\end{figure}

In this paper, we present a new strategy for non-convex optimization algorithms to escape local minima, inspired by time-varying cost landscapes.
Given a Hamiltonian, we construct an alternate Hamiltonian which shares minima with the original Hamiltonian only when the energy according to the original Hamiltonian is below a certain threshold.
An optimization algorithm repeatedly applied to each of the two Hamiltonians in sequence will escape any local minima with energy above this threshold.
The convergence of our technique only depends on the shape of the landscape; as a result, it can be used with both gradient-based and gradient-free optimization algorithms (including genetic optimization and simulated annealing).
We sketch the intuition behind our strategy in  Fig.~\ref{cartoon}.

We first discuss how to construct an alternate Hamiltonian from a given Hamiltonian. 
We then illustrate our \emph{Hamiltonian alternation} strategy with an application to a simple one-dimensional non-convex optimization problem.
Finally, we demonstrate the success of our strategy compared to multi-start optimization when minimizing the energy of the Sherrington-Kirkpatrick model of a spin glass~\cite{Sherrington_Kirkpatrick_1975, Mezard_Parisi_Virasoro_1987}. 
Constructing an alternate Hamiltonian is straightforward when the ground state energy is approximately known; regardless, one can use Hamiltonian alternation even when the ground state energy is unknown. 

\section{Theory of Hamiltonian Alternation}
Consider a Hamiltonian $H$ encoding an optimization problem of interest. 
We construct an alternate Hamiltonian:
\begin{equation}
    \widetilde{H} = K(H)~,
\end{equation}
composing $H$ with a scalar function $K$. 
If both the original Hamiltonian $H$ and alternate Hamiltonian $\widetilde{H}$ are smooth, we can take derivatives to understand the nature of their critical points.
Since
\begin{align}\label{altergrad}
    \partial_{i}\widetilde{H}(x)= K'(H(x))\partial_{i}H(x)~,
\end{align}
every critical point of $H$ is also a critical point of $\widetilde{H}$. 
At each of these points,
\begin{align}\label{Hess}
        \partial_{i}\partial_{j}\widetilde{H}(x)&=K''(H(x))\partial_{i}H\cdot\partial_{j}H+K'(H(x))\partial_{i}\partial_{j}H(x)\nonumber \\
        &=K'(H(x))\partial_{i}\partial_{j}H(x)~,
\end{align}
since the first derivatives of $H$ vanish.
The Hessian matrix $\partial_{i}\partial_{j}H$ is positive definite at each minimum of $H$; as a consequence, the sign of $K'(H)$ determines whether a minimum of $H$ is a minimum or maximum of $\widetilde{H}$. 
If one judiciously chooses $K$ such that $K'(H)$ is positive near the ground state energy, then the only minima that $H$ and $\widetilde{H}$ share are the ground states of $H$.
Optimizing over $H$ and $\widetilde{H}$ in sequence comprises a \emph{round} of Hamiltonian alternation.

Optimization algorithms can also have trouble escaping \emph{saddle points}. 
The more convex orthogonal directions of a saddle point, the harder it is to escape \cite{Hofman2018}.
In high-dimensional spaces, random matrix theory suggests that saddle points are exponentially more likely to occur than local minima \cite{Bengio_2014}. 
Any saddle point of $H$ is also a saddle point of $\widetilde{H}$, but the nature of the saddle depends on $K$.
Suppose $K'(H)$ is positive only in the neighborhood of the ground state.
At all other critical points, positive and negative entries of the Hessian are interchanged. 
The number of positive/negative entries of the Hessian determines the number of convex/concave directions of the saddle.
Therefore, a hard-to-escape saddle point of $H$ is easy to escape in $\widetilde{H}$ (and vice versa). 

This analysis of critical points extends to discrete optimization problems, as long as $H$ does not vary drastically between adjacent points in its state space. An instance of such a discrete optimization problem is the Sherrington-Kirkpatrick model discussed in the Applications section.

When the ground state energy can be approximated, one can choose many $K$ such that the global minimum of $H$ is the only shared minimum of $H$ and $\widetilde{H} = K(H)$.
One simple choice of $K$ is a quadratic function 
\begin{equation}\label{eq:quad}
 \widetilde{H} = H\cdot(2b -H)~,
\end{equation}
where $b$ lies above the ground state energy and below the energy of the next-lowest minima.

If the ground state energy is not known, one can use information obtained from previous attempts at minimizing $H$ to construct the alternate Hamiltonian. For example, repeatedly updating $b$ to the lowest value known so far allows an optimization algorithm with Hamiltonian alternation to improve its performance.  
We discuss this further in the Choosing Alternation Parameters section.

Simulated annealing always converges to the global minimum with a long enough annealing schedule \cite{Granville_Krivanek_Rasson_1994}.
Under mild assumptions, Hamiltonian alternation behaves similarly with sufficiently many rounds.
Choose an optimization algorithm that computes the nearest local minimum from its initial point, e.g. gradient descent with a small enough learning rate.
For any bounded region $U$ of the search space, Hamiltonian alternation can find the minimum of $H$ whenever $H$ and $\widetilde{H}$ share at least one minimum.
This is because the procedure visits critical points within $U$ until it reaches a shared minimum of $H$ and $\widetilde{H}$. 

\subsection{One-dimensional Example}\label{subsec:1d}

\begin{figure}[htb!]
    \centering
    \includegraphics[width=\linewidth]{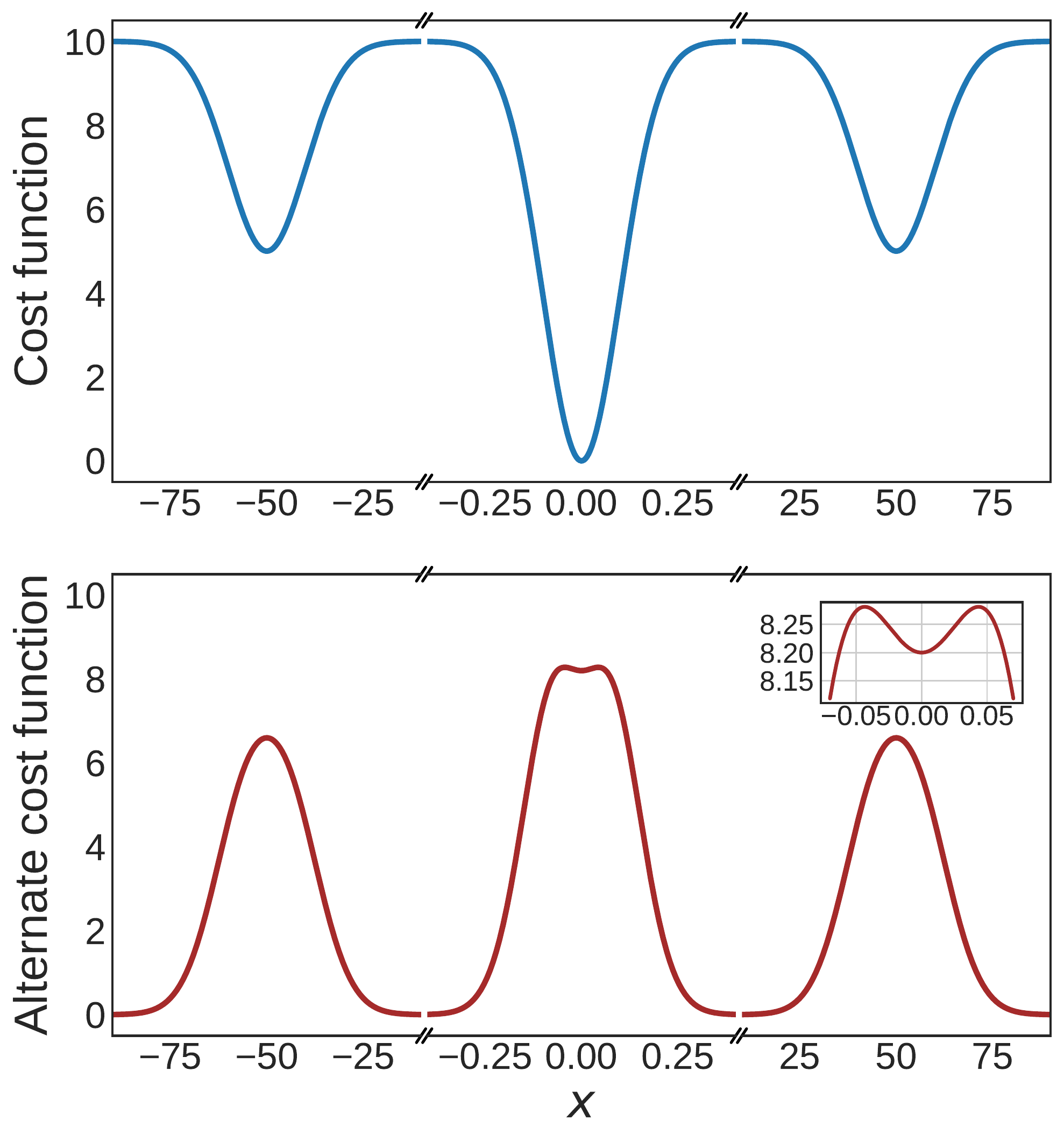}
    \caption{\footnotesize Hamiltonian alternation applied to a one-dimensional optimization problem. The original cost function has a narrow minimum at $x=0$, and the alternate cost function shares a minimum only at this position.}
    \label{fig:1d-function}
\end{figure}

We illustrate the technique of Hamiltonian alternation by applying it to a simple one-dimensional optimization problem.
In this problem, we minimize a sum of three Gaussian functions. The cost function has a narrow global minimum of value $0$ at $x=0$, flanked on either side by wide local minima of value $5$. See the illustration in Fig.~\ref{fig:1d-function}(a). 

In high-dimensional problems, the global minimum's basin of attraction is typically quite narrow relative to the size of the state space; our example mimics this behavior.
We construct an alternate cost function by choosing $b=0.9$ in Eq.~\ref{eq:quad}, as shown in Fig.~\ref{fig:1d-function}(b).
Notice that $b$ lies between the values of the two minima; as a result, the local minima are local maxima of the alternate cost function.
However, $x=0$ remains a local minimum of the alternate cost function, as shown in the inset of Fig.~\ref{fig:1d-function}(b). 

We solve this optimization problem with the Broyden-Fletcher-Goldfarb-Shanno (BFGS) algorithm.
BFGS is an iterative method for solving unconstrained nonlinear optimization problems using curvature information to precondition the gradient \cite{Nocedal_Wright_2006}.
However, since the global minimum has such a narrow attraction basin, randomly initialized BFGS rarely finds it.
To avoid this, a \emph{multi-start} strategy runs many copies of BFGS, each from a different initial point. This increases the chance of finding the global minimum.

We also implement BFGS with Hamiltonian alternation.
In one round of alternation, we let BFGS settle into a candidate minimum; then, after a slight perturbation, we re-apply BFGS to the alternate cost function.

\begin{figure}[ht!]
    \centering
    \includegraphics[width=\linewidth]{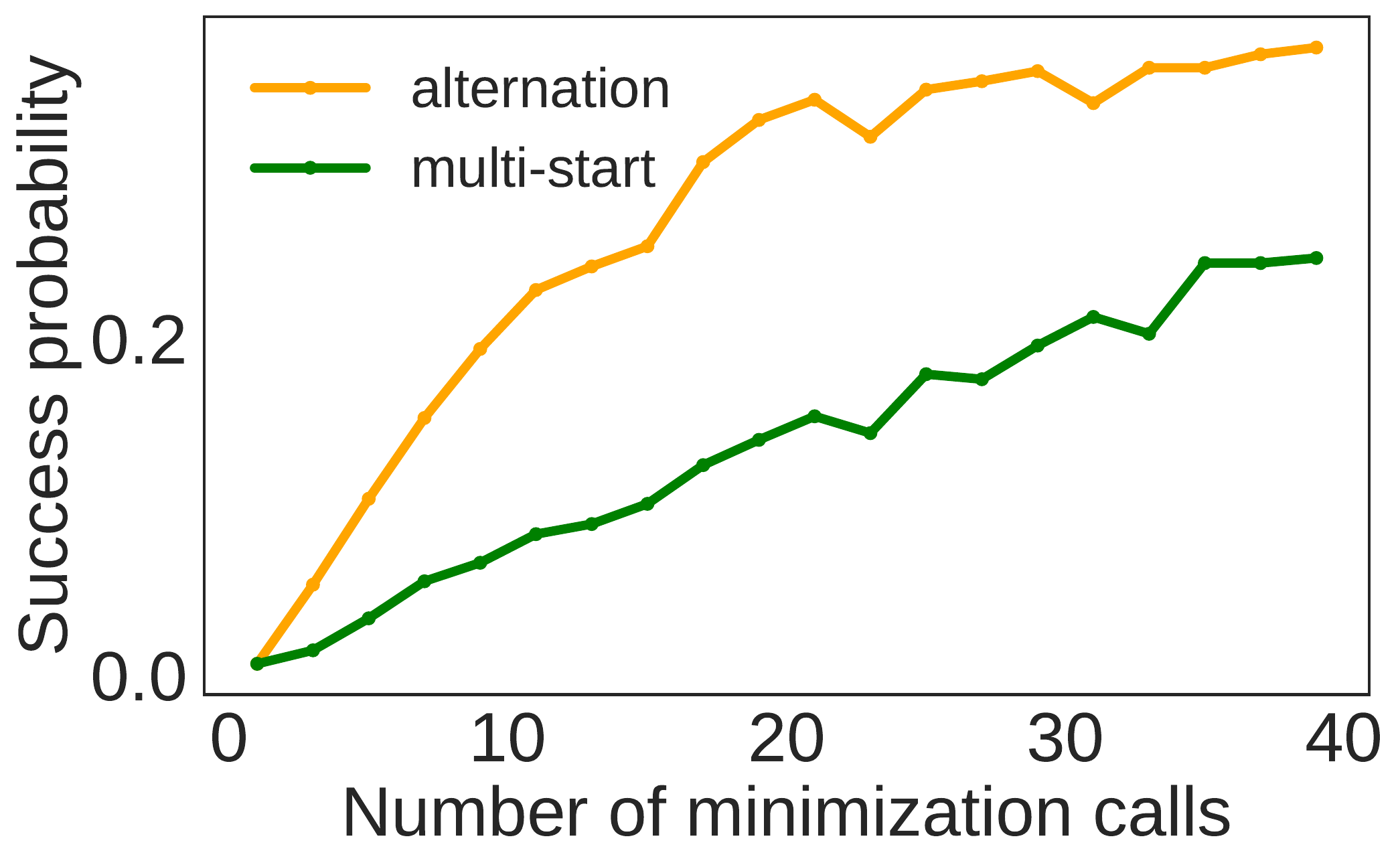}
    \caption{\footnotesize Comparison of BFGS with Hamiltonian alternation and BFGS with multi-start on a 1-dimensional problem, given a fixed number of minimization calls. In this example, alternation is the better strategy.}
        \label{fig:1d-success}
\end{figure}

We find that given a fixed number of minimization attempts, using many rounds of alternation outperforms a multi-start strategy. 
The probability of finding the minimum value varies slightly due to random initialization, so we compute the average across one hundred runs.
Fig.~\ref{fig:1d-success} compares the probability of finding the ground state between the two approaches given a fixed number of minimization calls.

\section{Choosing Alternation Parameters}\label{choice}
In order to use Hamiltonian alternation, one must choose the algorithm used to optimize the original and alternate cost functions.
Aside from this, we describe four important parameters that must be chosen before using our technique: \emph{alternation ratio} $a_r$, \emph{alternate cost function} $\widetilde{H} = K(H)$, \emph{switching rule}, and \emph{stopping criterion}. We walk through these choices one by one.

In our one-dimensional example, we compared two approaches:
$N$ randomly initialized calls to BFGS (\emph{multi-start}) and $N$ sequential calls to BFGS using $N/2$ rounds of Hamiltonian alternation.
One can also implement a hybrid of the two approaches that uses the same number of calls to the underlying algorithm.
We define the fraction of optimization algorithm calls dedicated to sequential Hamiltonian alternation as the \emph{alternation ratio} $a_r$, where $0 \leq a_r \leq 1$. The optimal choice of alternation ratio depends on the landscape of the optimization problem.
We hypothesize that increasing the value of $a_r$ leads to better performance when the basin of attraction for the global minimum is narrow.

The \emph{alternate cost function} is chosen by specifying the scalar function $K$ that transforms the original cost function.
This transformation should preserve known symmetries of the underlying problem.
It should also be chosen so that the basin associated with the global minimum does not shrink too much; otherwise, the algorithm may ``jump over'' the minimum.
For example, if the location of the global minimum in the alternate cost function is a small valley-shaped region nestled between high-valued maxima, an algorithm may have trouble finding it. 

One can also update the alternate cost function after each round.
This is especially helpful when the ground state energy is not known.
For example, when using the form of Eq.~\ref{eq:quad}, one can update the threshold $b$ with the lowest energy obtained thus far.
This ensures that alternation can reach lower-lying minima, although there is no guarantee of reaching the ground state.

We must also decide on a \emph{switching rule}.
This rule determines when an optimization algorithm (like BFGS) stops minimizing over a cost function and starts minimizing over the other cost function.
Possible criteria include a maximum number of iterations, or a minimum amount of improvement in the objective.
A natural choice is to allow the algorithm to reach its internal stopping criteria before switching the cost function.
However, one could use information obtained during previous rounds of alternation to implement a ``smart'' switching strategy.

Finally, we must choose a \emph{stopping criterion} that signals when our algorithm is finished. 
However, if the ground state energy is underestimated, then \emph{all} minima of the original cost function are maxima of the alternate cost function. 
In this case, Hamiltonian alternation will not terminate.
Some other stopping criterion, like a maximum number of rounds, can prevent Hamiltonian alternation from running indefinitely.

We summarize the steps of using Hamiltonian alternation with the pseudocode in Fig.~\ref{fig:alg}.

\begin{figure}[htb!]
\begin{algorithm}[H]
\SetAlgoLined
\KwData{cost function $H$   }
\KwResult{minimum value attained}
choose optimization algorithm\;
choose $a_r$, switching rule, stopping criterion\;
choose $K$ and construct $\widetilde{H}$; \\
\For{each initial point}{\While{stopping criterion not met}{
minimize $H$\;
optionally update $\widetilde{H}$\;
minimize $\widetilde{H}$\;
}}
\end{algorithm}
\caption{\footnotesize Pseudocode describing the Hamiltonian alternation metaheuristic.}
\label{fig:alg}
\end{figure}

\section{Applications}
The Hamiltonian alternation technique can be applied to both gradient-based and gradient-free optimization algorithms.
We demonstrate the success of our strategy by applying it to the problem of finding the ground state energy of a spin glass.
Spin glasses are an important object of study in condensed-matter physics \cite{Anderson_1988}.
Furthermore, many optimization problems can be recast as spin glasses \cite{Lucas_2014}.  

\subsection{Sherrington-Kirkpatrick Model of Spin Glass}
Following the work of Edwards and Anderson \cite{Edwards_1975}, Sherrington and Kirkpatrick (SK) developed a model of spin glass capturing the competition between quenched ferromagnetic and anti-ferromagnetic interactions \cite{Sherrington_Kirkpatrick_1975}.
This model has found applications in fields beyond physics, including biology and the study of neural networks \cite{Anderson_1990}. 

Consider a system with $N$ spins which can point either up or down. 
The SK model of spin glass is described by a Hamiltonian
\begin{equation}\label{SK_Ham}
    H_{SK}(\sigma) = - \frac{1}{\sqrt{N}} \sum_{1 \leq i < j \leq N} J_{ij} \sigma_{i}\sigma_{j}~,
\end{equation}
where the random couplings $J_{ij}$ are independently drawn from a standard Gaussian distribution.
The $1/\sqrt{N}$ factor ensures that the free energy per spin does not scale with $N$.
Finding the ground state of $H_{SK}$ is difficult because the size of the state space grows exponentially in $N$;
in fact, the decision problem variant is NP-complete \cite{Sherrington_2010}. 
The energy landscape has a large number of local minima \cite{Malsagov_2016}, therefore this problem is particularly challenging for many optimization algorithms.
As $N \xrightarrow{} \infty$, the ground state energy per spin almost always approaches a constant, which is approximately equal to $-0.7632$ \cite{Parisi_1980}. 

One method to find the ground state energy starts from a random spin configuration, and flips any spins that reduce the energy of the configuration.
The procedure repeats until the state cannot be improved with any single spin flip.
This strategy, called ``local search'', resembles gradient descent (which proceeds until no direction has a negative gradient).
In fact, local search can be used to characterize the energy spectra of this spin glass model \cite{Malsagov_2016}.

\begin{figure}[htb!]
    \includegraphics[width = \linewidth]{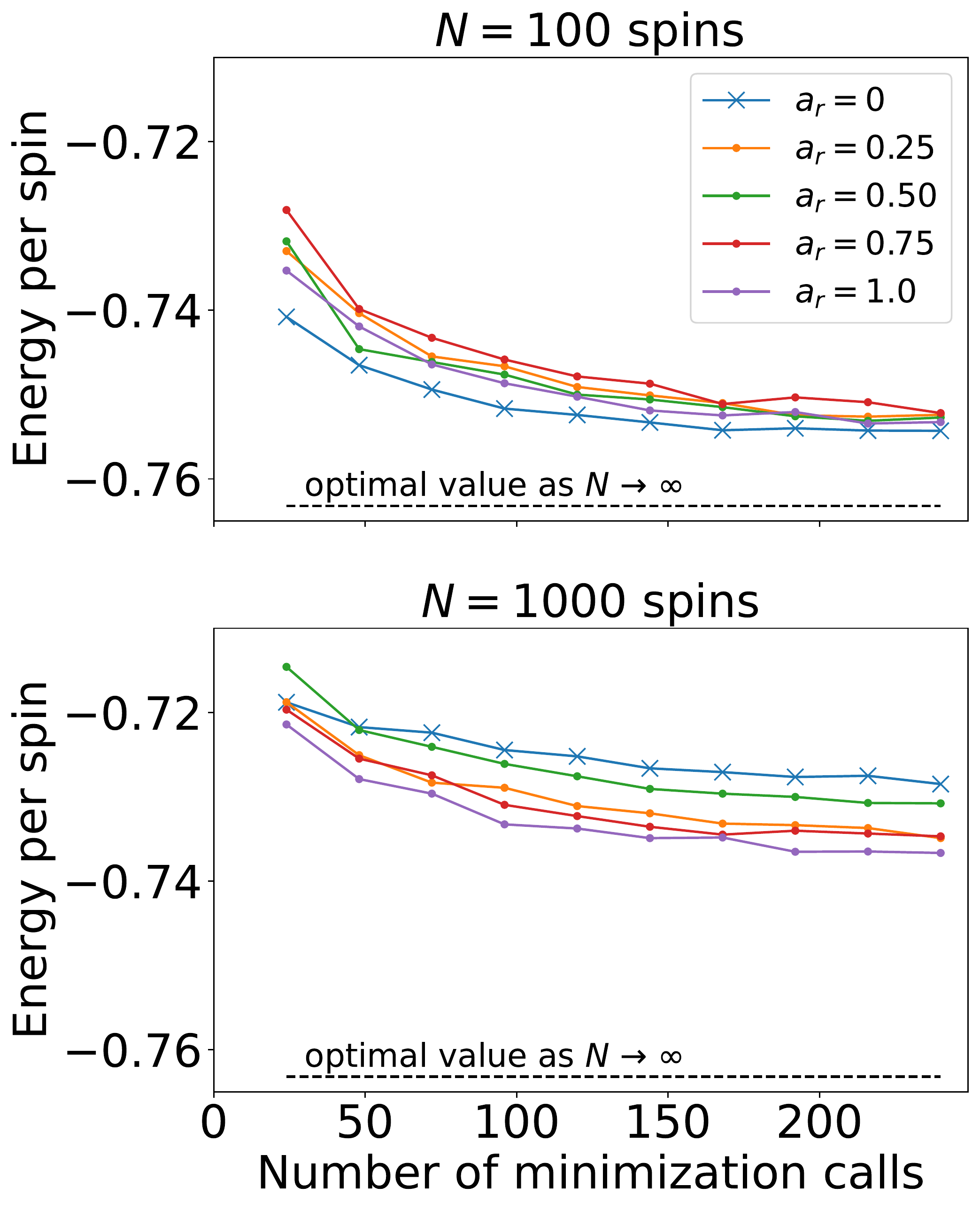}
    \caption{\footnotesize The minimum value of energy per spin, computed on instances of the Sherrington-Kirkpatrick Hamiltonian for $N=100$ spins (top) and $N=1000$ spins (bottom). 
    When $N=100$, the pure multi-start strategy (alternation ratio $a_r = 0$) performs best, but there is little variation in performance with increasing amounts of Hamiltonian alternation (i.e., increasing  $a_r$).
    At $N=1000$, Hamiltonian alternation systematically improves on a pure multi-start strategy.}
    \label{fig:sk}
\end{figure}

We minimize the SK Hamiltonian (Eq.~\ref{SK_Ham}) using local search with Hamiltonian alternation.
Since the exact ground state energy for a given instance is not known, we update the alternate Hamiltonian after each round.
In particular, we use the form of alternate Hamiltonian in Eq.~\ref{eq:quad}, choosing $b$ to be one percent less than the lowest energy value seen so far.
We apply this to instances with $N=100$ and $N=1000$ spins, varying the alternation ratio and total number of minimization calls.
We compute the energy per spin averaged across one hundred runs.
When $N=1000$, our results confirm that Hamiltonian alternation improves on a pure multi-start approach. See Fig.~\ref{fig:sk}(a).

The state space of an optimization problem grows with the number of input variables. 
For larger instances, the relative size of the ground state's basin of attraction is smaller.
Consequently, we expect Hamiltonian alternation to provide advantage only for large instances.
We find that SK instances with $N=100$ spins are too small for Hamiltonian alternation to provide advantage, although the performance does not vary much with alternation ratio.
Fig.~\ref{fig:sk}(b) illustrates this behavior.

\subsection{Limitations}
On small problems, Hamiltonian alternation is less likely to lead to improved performance when compared with a multi-start strategy.
We observe this for the SK model of spin glass at $N=100$ spins.
In general, we cannot predict exactly when Hamiltonian alternation improves the performance of an optimization algorithm; it may depend on the type of problem, size of problem, and choice of underlying algorithm.
We suggest that Hamiltonian alternation is most useful in high-dimensional settings and other problems where the global minima is challenging to find with a multi-start strategy.

Even if the problem is suitable for Hamiltonian alternation, it may be difficult to appropriately choose parameters like \emph{alternation ratio} and \emph{alternate cost function}. 
This issue is common among most algorithms for non-convex optimization.
Typically, there are no recipes for choosing an algorithm's parameters and predicting the impact on its performance \cite{Bergstra_2012}.
For example, a high learning rate in machine learning algorithms makes an algorithm more likely to skip over minima; in contrast, a low learning rate increases convergence time, and may cause the algorithm to get stuck in an undesirable local minimum \cite{Buduma_2017}.

\section{Discussion}
We have described a new strategy to help non-convex optimization algorithms escape from local minima. Alternating between the original cost function and a second cost function, computed easily from the first, allows an algorithm to find lower-energy solutions. 
We expect our method to be especially useful when low-lying minima have narrow attraction basins that cannot be found simply by running an algorithm from many randomly initialized points.

Our technique can be used with any optimization algorithm that may get stuck in local minima.
It is an open question whether our method improves algorithms that mix local steps with non-local exploration, e.g. a genetic algorithm with increasing amounts of recombination.
It is also unknown whether the \emph{switching rule} can be chosen synergistically with other algorithm parameters, like the cooling schedule in simulated annealing.

Optimization landscapes for variational quantum algorithms are almost always non-convex \cite{Huembeli2021, Cerezo2021}.
Hamiltonian alternation may enable variational algorithms to find lower-energy solutions.
It may be challenging to choose an alternate cost function that can be implemented with a quantum computer; however, multi-qubit quantum gates can be implemented with similar fidelity as two-qubit gates on some platforms \cite{gokhale2021quantum}.
In hybrid quantum-classical approaches that can compute the gradient with automatic differentiation \cite{Killoran2018, Killoran2021}, one can directly use the derivative of the alternate Hamiltonian for gradient descent.

\section*{Acknowledgments}
We thank Lorenzo Orecchia and Fred Chong for valuable discussions. AA is supported by the Yoichiro Nambu Graduate Fellowship courtesy of Department of Physics, University of Chicago. The work of KM is supported by the National Science Foundation Graduate Research Fellowship Program under Grant No. DGE-1746045. AM thanks the Simons Foundation for support. This work was primarily supported by the University of Chicago Materials Research Science and Engineering Center, which is funded by National Science Foundation under award number DMR-2011854. Any opinions, findings, and conclusions or recommendations expressed in this material are those of the author(s) and do not necessarily reflect the views of the National Science Foundation.

\bibliographystyle{apsrev4-1}
\bibliography{references}

\end{document}